\documentclass[11pt,a4paper]{article}%
\usepackage{amsfonts}
\usepackage{amsmath}
\usepackage{amssymb}
\usepackage[margin=1in,a4paper]{geometry}
\usepackage[sc,center,compact,noindentafter,medium]{titlesec}
\usepackage[onehalfspacing,nodisplayskipstretch]{setspace}
\usepackage{graphicx}
\usepackage{color}%
\definecolor{darkblue}{rgb}{.2, 0.2,.8}
\definecolor{darkgreen}{rgb}{0,0.5,0.3}
\definecolor{darkred}{rgb}{.8, .1,.1}

\providecommand{\U}[1]{\protect\rule{.1in}{.1in}}

\newtheorem{theorem}{\normalfont\scshape Theorem}[section]

\newtheorem{lemma}{\normalfont\scshape Lemma}[section]

\newtheorem{remark}{\normalfont\scshape Remark}[section]

\expandafter
\def \expandafter \normalsize \expandafter{\normalsize \setlength \abovedisplayskip{10pt plus 2pt minus 7pt}}
\expandafter
\def \expandafter \normalsize \expandafter{\normalsize \setlength \abovedisplayshortskip{0pt plus 2pt}}
\expandafter
\def \expandafter \normalsize \expandafter{\normalsize \setlength \belowdisplayskip{10pt plus 2pt minus 7pt}}
\expandafter
\def \expandafter \normalsize \expandafter{\normalsize \setlength \belowdisplayshortskip{5pt plus 2pt minus 3pt}}

\numberwithin{equation}{section}
\begin{document}

\title{ }

\begin{center}
{\LARGE \textsc{asymptotics for the generalized autoregressive conditional
duration model}}%

\renewcommand{\thefootnote}{}
\footnote{
\hspace{-7.2mm}
$^{a}%
$Department of Economics, University of Bologna, Italy and Department of Economics, University of Exeter, UK.
\newline$^{b}%
$Department of Mathematical Sciences, University of Copenhagen, Denmark.
\newline$^{c}$Department of Economics, University of Copenhagen, Denmark.
\newline
We thank Matias Cattaneo and seminar participants at Princeton University for comments and suggestions.
A.~Rahbek and G.~Cavaliere gratefully acknowledge support from the Independent Research Fund Denmark (DFF Grant 7015-00028) and
the Italian Ministry of University and Research (PRIN 2020 Grant 2020B2AKFW). T.~Mikosch's research is partially supported by the Independent Research Fund Denmark (DFF Grant 9040-00086B).
Correspondence to: Giuseppe Cavaliere, Department of
Economics, University of Bologna, email giuseppe.cavaliere@unibo.it.
}
\addtocounter{footnote}{-1}
\renewcommand{\thefootnote}{\arabic{footnote}}%
{\normalsize \vspace{0.1cm} }

{\large \textsc{Giuseppe Cavaliere}}$^{a}${\large \textsc{, Thomas Mikosch}%
}$^{b}$, {\large \textsc{Anders Rahbek}}$^{c}$

{\large \textsc{and Frederik Vilandt}}$^{c}${\normalsize \vspace{0.2cm}%
\vspace{0.2cm}}

May 25, 2023{\normalsize \vspace{0.2cm}\vspace{0.2cm}}

Abstract\vspace{-0.15cm}
\end{center}

Engle and Russell (1998, \emph{Econometrica}, 66:1127--1162) apply results
from the GARCH\ literature to prove consistency and asymptotic normality of
the (exponential) QMLE\ for the generalized autoregressive conditional
duration (ACD)\ model, the so-called ACD(1,1), under the assumption of strict
stationarity and ergodicity. The GARCH results, however, do not account for
the fact that the number of durations over a given observation period is
random. Thus, in contrast with Engle and Russell (1998), we show that strict
stationarity and ergodicity alone are not sufficient for consistency and
asymptotic normality, and provide additional sufficient conditions to account
for the random number of durations. In particular, we argue that the durations
need to satisfy the stronger requirement that they have finite mean. \bigskip


\section{Introduction}

\label{sec intro}\textsc{In the seminal paper }by Engle and Russell (1998, ER
henceforth), autoregressive conditional duration (ACD) models were introduced
for modeling durations between financial transactions. Given some observation
period $[0,T]$, say, with $n(T)$ observed event times $\left\{  t_{i}\right\}
_{i=1}^{n(T)}$, $0<t_{1}<t_{2}<\cdots<t_{n(T)}\leq T$, the durations $x_{i}$
are given by $x_{i}=t_{i}-t_{i-1}$ and modeled as
\begin{align}
x_{i}  &  =\psi_{i}\left(  \theta\right)  \varepsilon_{i},\text{
\ \ }i=1,\ldots,n(T)\,,\label{eq ACD_1}\\
\psi_{i}\left(  \theta\right)   &  =\omega+\alpha x_{i-1}+\beta\psi
_{i-1}\left(  \theta\right)  \label{eq ACD_2}%
\end{align}
where the innovations $\{\varepsilon_{i}\}$ are i.i.d., strictly positive,
with unit mean, $\mathbb{E}[\varepsilon_{i}]=1$. The quasi\ maximum likelihood
estimator (QMLE) of $\theta=(\omega,\alpha,\beta)^{\prime}\in\Theta
\subset\mathbb{R}^{3}$ is defined as $\hat{\theta}_{T}=\arg\max_{\theta
\in\Theta}L_{n(T)}\left(  \theta\right)  $, with $L_{n(T)}\left(
\theta\right)  $ the exponential likelihood,%
\begin{equation}
L_{n(T)}\left(  \theta\right)  =-\sum_{i=1}^{n(T)}\ell_{i}(\theta)\text{,
}\ell_{i}(\theta)=\log\psi_{i}\left(  \theta\right)  +\frac{x_{i}}{\psi
_{i}\left(  \theta\right)  }\,,\qquad T\geq0\,, \label{eq: EACD likelihood}%
\end{equation}
with $\omega,\alpha$ and $\beta$ positive, and initial values $(x_{0},\psi
_{0}(\theta))^{\prime}=\gamma$.

ER\ note that the likelihood function in (\ref{eq: EACD likelihood}) is
identical to the likelihood function of the GARCH(1,1) model with Gaussian
innovations. Hence, for their main result (p. 1135), ER\ refer to Lee and
Hansen (1994) to conclude that under strict stationarity and ergodicity of the
durations $x_{i}$, $\hat{\theta}_{T}$ is consistent and asymptotically normal
at the usual rate; importantly, their conditions allow for $\alpha+\beta\geq
1$, and hence for durations with no finite mean. In contrast, using a new
lemma which extends the arguments in Lee and Hansen (1994) to the ACD case, we
argue that the additional condition $\alpha+\beta<1$, which implies a finite
mean, is sufficient.

To establish asymptotic normality, standard arguments require that the
(normalized) score and information, evaluated at the true value $\theta
=\theta_{0}$:%
\begin{align}
S_{n(T)}  &  =\left.  \tfrac{\partial L_{n(T)}(\theta)}{\partial\theta
}\right\vert _{\theta=\theta_{0}}=n(T)^{-1/2}\sum_{i=1}^{n(T)}\xi_{i}%
\,,\quad\xi_{i}=\left.  \tfrac{\partial\ell_{i}(\theta)}{\partial\theta
}\right\vert _{\theta=\theta_{0}}\label{eq S}\\
I_{n(T)}  &  =\left.  -\tfrac{\partial^{2}L_{n(T)}(\theta)}{\partial
\theta\partial\theta^{\prime}}\right\vert _{\theta=\theta_{0}}=n(T)^{-1}%
\sum_{i=1}^{n(T)}\zeta_{i}\,,\qquad\zeta_{i}=\left.  -\tfrac{\partial^{2}%
\ell_{i}(\theta)}{\partial\theta\partial\theta^{\prime}}\right\vert
_{\theta=\theta_{0}} \label{eq I}%
\end{align}
satisfy a central limit theorem (CLT)\ and a law of large numbers (LLN),
respectively. The ACD setting, however, is not standard as the number of
observations $n(T)$ is \emph{random} and not independent of the sequences
$\{\xi_{i}\}$ and $\{\zeta_{i}\}$. Note in this respect that the fact that the
CLT\ and the LLN\ hold for the case of a deterministic number $n$ of
observations, that is
\begin{align}
S_{n}  &  =n^{-1/2}\sum_{i=1}^{n}\xi_{i}\overset{d}{\rightarrow}N\left(
0,\Omega_{S}\right)  \text{,}\,\label{eq CLT for deterministic n}\\
I_{n}  &  =n^{-1}\sum_{i=1}^{n}\zeta_{i}\,\overset{\text{a.s.}}{\rightarrow
}\Omega_{I}\text{,} \label{eq LLN for deterministic n}%
\end{align}
does not imply that their random $n(T)$ analogues in\textbf{ }(\ref{eq S}%
)-(\ref{eq I}) hold. Therefore, arguments based on Lee and Hansen (1994) as in
ER, which assume $n$ deterministic, do not apply directly.

In this note, by using a new Lemma which extends Lee and Hansen (1994)\ to the
case of a random number of observations, we show that under the additional
more restrictive\textbf{ }condition $\alpha_{0}+\beta_{0}<1$ (or,
equivalently, that the durations have finite unconditional expectation) the
QMLE\ is consistent and asymptotically normal at the standard $\sqrt{T}$ rate.

The key additional condition needed is that the random number of durations
$n(T)$ satisfies
\begin{equation}
n(T)/T\overset{\text{a.s.}}{\rightarrow}1/\mu\text{ , with }\mu=\mathbb{E}%
\left(  x_{i}\right)  \text{.} \label{eq N_T/T convergence}%
\end{equation}
This in turn (as $n(T)\rightarrow\infty$ a.s.) is sufficient for the
deterministic $n$ LLN in (\ref{eq LLN for deterministic n}) to imply that its
random $n$ analogue (\ref{eq I}) holds. To establish the random $n$ CLT
in\textbf{ }(\ref{eq S}), we replace the deterministic $n$ CLT\ in
(\ref{eq CLT for deterministic n}) with its stronger functional version
\[
S_{n}\left(  \cdot\right)  =n^{-1/2}\sum_{i=1}^{\left\lfloor n\cdot
\right\rfloor }\xi_{i}\overset{d}{\rightarrow}\Omega_{S}^{1/2}B\left(
\cdot\right)  \text{,}\,
\]
where $B$ is a standard multivariate Brownian motion.

We finally notice that when $\alpha_{0}+\beta_{0}>1$, which implies ergodicity
provided also $E[\ln\left(  \alpha_{0}\varepsilon_{i}+\beta_{0}\right)  ]<0$
holds, asymptotically normality is no longer guaranteed; results in Cavaliere,
Mikosch, Rahbek and Vilandt (2022) for the simple ACD\ model with $\beta=0$,
suggest that $\sqrt{T}$ asymptotic normality indeed breaks down in this case.

\section{Main result}

In order to derive the asymptotic distribution of the QMLE\ for the ACD model
given by (\ref{eq ACD_1})-(\ref{eq ACD_2}) we first introduce the following
general lemma, which extends the results in Lee and Hansen (1994) to allow for
a random number of observations.

\begin{lemma}
\label{new: main lemma}Consider $Q_{n}(\varphi)\in\mathbb{R}$, which is a
random function of the deterministic sample size $n$ and the parameter
$\varphi\in\Phi\subseteq\mathbb{R}^{k}$. Assume that $Q_{n}(\cdot
):\mathbb{R}^{k}\rightarrow\mathbb{R}$ is three times continuously
differentiable\ in $\varphi$, and that for $\varphi_{0}$ in the interior of
$\Phi$ it holds that as $n\rightarrow\infty$:$\smallskip$

\noindent$%
\begin{tabular}
[c]{ll}%
(C.1) & $n^{-1/2}\partial Q_{[n\cdot]}(\varphi_{0})/\partial\varphi\overset
{w}{\rightarrow}\Omega_{S}^{1/2}B(\cdot)$, $\ \Omega_{S}>0$,\\
(C.2) & $-n^{-1}\partial^{2}Q_{n}(\varphi_{0})/\partial\varphi\partial
\varphi^{\prime}\overset{\text{a.s}.}{\rightarrow}\Omega_{I}>0$,\\
(C.3) & $\max_{h,i,j=1,...,k}\sup_{\varphi\in N(\varphi_{0})}\left\vert
n^{-1}\frac{\partial^{3}Q_{n}(\varphi)}{\partial\varphi_{h}\partial\varphi
_{i}\partial\varphi_{j}}\right\vert \leq\tau_{n}\rightarrow\tau$ a.s.,
\end{tabular}
\ \smallskip$

\noindent where $B\left(  \cdot\right)  $ is a $k$-dimensional Brownian
motion, $N(\varphi_{0})$ is a neighborhood of $\varphi_{0},$ and
$0<\tau<\infty$. Moreover, with $n\left(  t\right)  $, $t\geq0$, a counting
process defined on the same probability space as $Q_{n}\left(  \varphi\right)
$, assume that with $c\in(0,\infty)$ a constant:$\smallskip$

\noindent$%
\begin{tabular}
[c]{lc}%
(C.4) & As $T\rightarrow\infty$, $n\left(  T\right)  /T\overset{\text{a.s.}%
}{\rightarrow}c$.
\end{tabular}
\ \smallskip$

\noindent Consider next $Q_{n\left(  T\right)  }\left(  \varphi\right)  $
which is a random function of the random sample size $n\left(  T\right)  $ and
$\varphi\in\Phi\subseteq\mathbb{R}^{k}$. Then, as $T\rightarrow\infty$, with
probability tending to one, there exists a fixed open neighborhood
$U(\varphi_{0})\subseteq N(\varphi_{0}),$ $\varphi_{0}\in U(\varphi_{0})$,
such that:$\smallskip$

\noindent$%
\begin{tabular}
[c]{ll}%
(i) & There exists a maximum point $\hat{\varphi}_{T}$ of $Q_{n\left(
T\right)  }(\varphi)$ in $U(\varphi_{0})$ and $Q_{n(T)}(\varphi)$ is concave
in\\
& $U(\varphi_{0})$; in particular, $\hat{\varphi}_{T}$ is unique and solves
$\partial Q_{n\left(  T\right)  }(\hat{\varphi}_{T})/\partial\varphi=0$,\\
(ii) & $\hat{\varphi}_{T}\overset{p}{\rightarrow}\varphi_{0}$,\\
(iii) & $T^{1/2}(\hat{\varphi}_{T}-\varphi_{0})\overset{d}{\rightarrow
}N(0,\Sigma)$, $\Sigma=c\Omega_{I}^{-1}\Omega_{S}\Omega_{I}^{-1}$.
\end{tabular}
\ $\bigskip
\end{lemma}

The proof of Lemma \ref{new: main lemma} is given in the appendix. Note that
Assumption (C.4) can be replaced by $n\left(  T\right)  \rightarrow\infty$
a.s. and $n\left(  T\right)  /T\overset{p}{\rightarrow}c$,\thinspace
\thinspace$0<c<\infty\,$.

Our main result is as follows.

\begin{theorem}
\label{thm main}For the ACD\ model (\ref{eq ACD_1})-(\ref{eq ACD_2}) with true
parameter value denoted by $\theta_{0}$, if: (i) $\{\varepsilon_{i}\}$ is an
i.i.d. sequence of r.v.s with support $(0,\infty)$, pdf $f_{\varepsilon
}\left(  \cdot\right)  $ bounded away from zero on compact subsets of $\left(
0,\infty\right)  $, $\mathbb{E}[\varepsilon_{i}]=1$ and $\mathbb{E}%
[\varepsilon_{i}^{2}]<\infty$, (ii) $\alpha_{0}+\beta_{0}<1$, then with
$\theta_{0}$ an interior point, the maximizer of $L_{n(T)}\left(
\theta\right)  $ in (\ref{eq: EACD likelihood}) will be consistent and
asymptotically normal at the standard $\sqrt{T}$-rate, with a covariance
matrix given by $\left(  1/\mu\right)  \Omega_{I}^{-1}\Omega_{S}\Omega
_{I}^{-1}$. Here $\mu=\mathbb{E}\left(  x_{i}\right)  <\infty$, and
$\Omega_{I}=\mathbb{E[}\zeta_{i}]$, $\Omega_{S}=\mathbb{E[}\xi_{i}\xi
_{i}^{\prime}]$ are given by (\ref{eq LLN for deterministic n})\ and
(\ref{eq CLT for deterministic n}) respectively.
\end{theorem}

\begin{remark}
Theorem \ref{thm main} can be extended by replacing the i.i.d. condition (i)
with the milder assumption that $\{\varepsilon_{i}\}$ is strictly stationary
and ergodic with (conditional) mean one, see Lee and Hansen (1994).
\end{remark}

\section*{References}

\smallskip\noindent\textsc{Cavaliere, G., Mikosch, T., Rahbek A., and Vilandt,
F. }(2022) The econometrics of financial duration modeling. \textsc{arXiv}:2208.02098.

\smallskip\noindent\textsc{Engle, R.F. and Russell, J.R.}\ (1998)
Autoregressive conditional duration: a new model for irregularly spaced
transaction data. \emph{Econometrica}, 66:1127--1162.

\smallskip\noindent\textsc{Gut, A.}\ (2009) \emph{Stopped Random Walks: Limit
Theorems and Applications}. Springer, NY.

\smallskip\noindent\textsc{Jensen, S.T. and Rahbek, A.} (2004) Asymptotic
inference for nonstationary GARCH. \emph{Econometric Theory, }20:1203--1226.

\smallskip\noindent\textsc{Jensen, S.T. and Rahbek, A.} (2007) On the law of
large numbers for (geometrically) ergodic Markov chains. \emph{Econometric
Theory, }23:761--766.

\smallskip\noindent\textsc{Lee, S. and Hansen, B.}\ (1994) Asymptotic theory
for the \textrm{GARCH}$(1,1)$\ quasi-maximum likelihood estimator.
\emph{Econometric Theory,} 10:29--52.

\smallskip\noindent\textsc{Meitz, M. and Saikkonen, P.}\ (2008) Ergodicity,
mixing, and existence of moments of a class of Markov models with applications
to GARCH and ACD models. \emph{Econometric Theory,} 24:1291--1320.

\smallskip\noindent\textsc{Phillips, P.C.B. and Solo, V.} (1992) Asymptotics
for linear processes. \emph{The Annals of Statistics}, 20:971--1001.

\appendix

\section*{Appendix}

\section{Proofs}

\subsection{Proof of Lemma \ref{new: main lemma}:}

We first consider the asymptotic behaviour as $T\rightarrow\infty$ for the
score, the second order derivative and the third order derivatives of
$Q_{n(T)}\left(  \varphi\right)  $. Next, we use these results to establish
\emph{(i)--(iii)}.

\noindent\emph{Score: }It holds that with $\partial Q_{n\left(  T\right)
}=\left.  \partial Q_{n\left(  T\right)  }\left(  \varphi\right)
/\partial\varphi\right\vert _{\varphi=\varphi_{0}},$%
\begin{equation}
n(T)^{-1/2}\partial Q_{n(T)}\overset{d}{\rightarrow}N\left(  0,\Omega
_{S}\right)  \text{.} \label{in app: random score}%
\end{equation}
To see this, let $\partial Q_{[Tc]}=\left.  \partial Q_{[Tc]}\left(
\varphi\right)  /\partial\varphi\right\vert _{\varphi=\varphi_{0}},$ and
decompose $n(T)^{-1/2}\partial Q_{n(T)}$ as%
\[
n(T)^{-1/2}\partial Q_{n(T)}=a_{T}^{-1/2}([Tc]^{-1/2}\partial Q_{[Tc]}%
)+a_{T}^{-1/2}A_{T}\text{, }%
\]
where $A_{T}=[Tc]^{-1/2}(\partial Q_{n(T)}-\partial Q_{[Tc]})$ and
$a_{T}=\frac{n(T)}{[Tc]}$. By conditions (C.1) and (C.4), $[Tc]^{-1/2}\partial
Q_{[Tc]}\overset{d}{\rightarrow}N\left(  0,\Omega_{S}\right)  $ and
$a_{T}\rightarrow1$ (a.s.). It remains to show that $A_{T}=o_{p}\left(
1\right)  $. F{or any $M,\delta>0$},%
\[
\mathbb{P}\left(  \left\Vert A_{T}\right\Vert >M\right)  =\mathbb{P(}%
\left\Vert A_{T}\right\Vert >M,\text{ }|n(T)/T-c|>\delta)+\mathbb{P(}%
\left\Vert A_{T}\right\Vert >M,\text{ }|n(T)/T-c|\leq\delta)\text{.}%
\]
Here, $\mathbb{P(}\left\Vert A_{T}\right\Vert >M,$ $|n(T)/T-c|>\delta
)\leq\mathbb{P}\left(  |n(T)/T-c|>\delta\right)  \rightarrow0$ by (C.4).
Next,
\begin{align*}
\mathbb{P(}\left\Vert A_{T}\right\Vert  &  >M,\text{ }|n(T)/T-c|\leq
\delta)=\mathbb{P(}[Tc]^{-1/2}\left\Vert \partial Q_{n(T)}-\partial
Q_{[Tc]}\right\Vert >M,\text{ }|n(T)/T-c|\leq\delta)\\
&  \leq\mathbb{P}([Tc]^{-1/2}\max_{c-\delta\leq u\leq c+\delta}\left\Vert
\partial Q_{[Tu]}-\partial Q_{[Tc]}\right\Vert >M)\\
&  \leq2\,\mathbb{P}(\max_{u\leq\,\delta}\left\Vert [Tc]^{-1/2}\partial
Q_{[Tu]}\right\Vert >{M)}\rightarrow2\,\mathbb{P}\left(  \left.  \max
_{u\leq\delta}\left\Vert \Omega_{S}^{1/2}\emph{B}(u)\right\Vert \right\vert
>c^{1/2}M\right)  \,,
\end{align*}
{as $T\rightarrow\infty$. As }$\delta$ can be arbitrarily small, it follows
that $A_{T}=o_{p}\left(  1\right)  $.

\noindent\emph{Second order derivative: }Since (C.4) implies $n(T)\rightarrow
\infty$ a.s., then by Gut (2009, Theorem 2.1) it holds that (C.2) implies%
\begin{equation}
-n(T)^{-1}\partial^{2}Q_{n(T)}\left(  \varphi_{0}\right)  /\partial
\varphi\partial\varphi^{\prime}\rightarrow\Omega_{I}>0\text{ a.s.}
\label{in app: random information}%
\end{equation}

\noindent\emph{Third order derivatives: }By (C.3),
\begin{equation}
\max_{h,i,j=1,...,k}\sup_{\varphi\in N(\varphi_{0})}\left\vert n(T)^{-1}%
\frac{\partial^{3}Q_{n(T)}(\varphi)}{\partial\varphi_{h}\partial\varphi
_{i}\partial\varphi_{j}}\right\vert \leq\tau_{n(T)}
\label{in app:: random third}%
\end{equation}
and hence, since $\tau_{n}\rightarrow\tau$ a.s., by (C.4)\ and again using Gut
(2009, Theorem 2.1), $\tau_{n(T)}\rightarrow\tau$ a.s.

\bigskip

\noindent\emph{Establishing (i)--(iii): }These hold by using
(\ref{in app: random score})--(\ref{in app:: random third}) together with the
arguments in the proof of Lemma 1 in Jensen and Rahbek (2004), replacing
$T\,$there by $n(T)$, and setting $\ell_{T}\left(  \varphi\right)
=-n(T)^{-1}Q_{n(T)}\left(  \varphi\right)  $. Specifically,
(\ref{in app: random score}) replaces condition (A.1) in Jensen and Rahbek
(2004), (\ref{in app: random information}) replaces their condition (A.2), and
(\ref{in app:: random third}) replaces their condition (A.3). A detailed
derivation is provided next. 

Note first that by definition $Q_{n(T)}(\varphi)$ is continuous in $\varphi$
and hence attains its maximum in any compact neighborhood $K(\varphi
_{0},r)=\{\varphi|\ \Vert\varphi-\varphi_{0}\Vert\leq r\}\subseteq
N(\varphi_{0})$ of $\varphi_{0}$. With $v_{\varphi}=\varphi-\varphi_{0},$ and
$\varphi^{\ast}$ on the line from $\varphi$ to $\varphi_{0}$, Taylor's formula
gives%
\begin{equation}
Q_{n(T)}(\varphi)-Q_{n(T)}(\varphi_{0})=\partial Q_{n(T)}(\varphi
_{0})v_{\varphi}-\left(  -\tfrac{1}{2}v_{\varphi}^{\prime}\partial^{2}%
Q_{n(T)}(\varphi^{\ast})v_{\varphi}\right)  \text{,}\label{new: eq5}%
\end{equation}
where $\partial Q_{n(T)}(\varphi)=\partial Q_{n(T)}(\varphi)/\partial\varphi$
and $\partial^{2}Q_{n(T)}(\varphi)=\partial^{2}Q_{n(T)}(\varphi)/\partial
\varphi\partial\varphi^{\prime}.$ The second term on the rhs of
(\ref{new: eq5}), normalized by $n(T)$, can be expressed as%
\begin{equation}
v_{\varphi}^{\prime}\left[  \Omega_{I}+(-n(T)^{-1}\partial^{2}Q_{n(T)}%
(\varphi_{0})-\Omega_{I})-n(T)^{-1}(\partial^{2}Q_{n(T)}(\varphi^{\ast
})-\partial^{2}Q_{n(T)}(\varphi_{0}))\right]  v_{\varphi}\text{.}%
\label{new: reperat}%
\end{equation}
Denote by $\rho_{n(T)}$ and $\rho$, $\rho>0$, the smallest eigenvalues of
$\left[  \left(  -n(T)^{-1}\partial^{2}Q_{n(T)}(\varphi_{0})\right)
-\Omega_{I}\right]  $ and $\Omega_{I}$ respectively. By
(\ref{in app: random information}) and the fact that the smallest eigenvalue
of a $k\times k$ symmetric matrix $M$, $\inf_{\{v\in\mathbb{R}^{k}|\Vert
v\Vert=1\}}v^{\prime}Mv$ is continuous in $M$, $\rho_{n(T)}\rightarrow0$ a.s.
Using (\ref{new: eq5}), then (\ref{in app: random score}%
)--(\ref{in app:: random third}) imply that
\begin{align*}
\sup_{\varphi:v_{\varphi}=r}n(T)^{-1}[Q_{n(T)}(\varphi)-Q_{n(T)}(\varphi_{0})]
&  \leq\Vert n(T)^{-1}\partial Q_{n(T)}(\varphi_{0})\Vert r-\tfrac{1}%
{2}\left[  \rho+\rho_{n(T)}-\tilde{\tau}_{n(T)}r\right]  r^{2}\\
&  \overset{p}{\rightarrow}-\tfrac{1}{2}\left[  \rho-\tilde{\tau}r\right]
r^{2}\text{,}%
\end{align*}
where $\tilde{\tau}_{n(T)}=k^{3/2}\tau_{n(T)}$ and $\tilde{\tau}=k^{3/2}\tau$.
Therefore, if $r<\rho/\tilde{\tau}$, the probability that $Q_{n(T)}(\varphi)$
attains its maximum on the boundary of $K(\varphi_{0},r)$ tends to zero. Next,
for $\varphi\in K(\varphi_{0},r)$ and $v\in\mathbb{R}^{k}$, rewriting
$v^{\prime}\partial^{2}Q_{n(T)}(\varphi)v$ as in (\ref{new: reperat}),
$-n(T)^{-1}v^{\prime}\partial^{2}Q_{n(T)}(\varphi)v\geq\Vert v\Vert^{2}%
(\rho+\rho_{T}-r\tilde{\tau}_{n(T)})$ which tends in probability to $\Vert
v\Vert^{2}(\rho-r\tilde{\tau})$. Hence, if $r<\rho/\tilde{\tau}$ the
probability that $Q_{n(T)}(\varphi)$ is strongly concave in the interior of
$K(\varphi_{0},r)$ tends to $1$, and therefore it has at most one stationary point.

This establishes \emph{(i)}: If $r<\rho/\tilde{\tau}$ and $K(\varphi
,r)\subseteq N(\varphi_{0}),$ there is with probability tending to one exactly
one solution $\hat{\varphi}_{T}$ to the likelihood equation in the interior of
$K(\varphi,r)$, $U(\varphi_{0})$. It is the unique maximum point of
$Q_{n(T)}(\varphi)$ in $U(\varphi_{0})$ and, as it is a stationary point, it
solves $\partial Q_{n(T)}(\varphi)=0$.

To establish \emph{(ii), }note that by the same argument, for any $\delta$,
$0<\delta<r$ there is with a probability tending to one a solution to the
likelihood equation in $K(\varphi_{0},\delta)$. As $\hat{\varphi}_{T}$ is the
unique solution to the likelihood equation in $K(\varphi_{0},r)$, it must
therefore be in $K(\varphi_{0},\delta)$ with a probability tending to $1$.
Hence we have proved that $\hat{\varphi}_{T}$ is consistent. That is, for any
$0<\delta<r$, the probability that $\hat{\varphi}_{T},$ $\Vert\hat{\varphi
}_{T}-\varphi_{0}\Vert\leq\delta,${\ is\ a\ unique\ solution to }$\partial
Q_{n(T)}(\varphi)=0$ in $K(\varphi_{0},r)$ tends to one, as desired.

Turning to \emph{(iii)}: By the result in (\ref{in app: random score}) and by
Taylor's formula to $\partial Q_{n(T)}(\varphi)/\partial\varphi_{j}$%
\begin{equation}
n(T)^{-1/2}\partial Q_{n(T)}(\varphi_{0})=\left(  n(T)/T\right)  ^{1/2}%
(\Omega_{I}+A_{T}(\hat{\varphi}_{T}))T^{1/2}(\hat{\varphi}_{T}-\varphi
_{0}),\label{new: eq7}%
\end{equation}
for $j=1,\ldots,k$. Here the elements in the matrix $A_{T}(\hat{\varphi}_{T})$
are of the form $v_{1}^{\prime}(-n(T)^{-1}\partial^{2}Q_{n(T)}(\varphi
_{T}^{\ast})-\Omega_{I})v_{2}$ with $v_{1},v_{2}$ unit vectors in
$\mathbb{R}^{k}$ and $\varphi_{T}^{\ast}$ a point on the line from
$\varphi_{0}$ to $\hat{\varphi}_{T}$ (where $\varphi_{T}^{\ast}$ depends on
the first vector $v_{1}$). Next, for any vectors $v_{1},v_{2}\in\mathbb{R}%
^{k},$ and any $\varphi\in N(\varphi_{0})$, using (\ref{in app:: random third}%
),
\begin{equation}
n(T)^{-1}\left\vert v_{1}^{\prime}\left(  \partial^{2}Q_{n(T)}(\varphi
)-\partial^{2}Q_{n(T)}(\varphi_{0})\right)  v_{2}\right\vert \leq\left\Vert
v_{1}\right\Vert \left\Vert v_{2}\right\Vert \left\Vert \varphi-\varphi
_{0}\right\Vert \tilde{\tau}_{n(T)}\text{.}\label{new: eq.4}%
\end{equation}
Using (\ref{new: eq.4}),
\[
|v_{1}^{\prime}(-n(T)^{-1}\partial^{2}Q_{n(T)}(\varphi_{T}^{\ast})-\Omega
_{I})v_{2}|\leq|v_{1}^{\prime}(-n(T)^{-1}\partial^{2}Q_{n(T)}(\varphi
_{0})-\Omega_{I})v_{2}|+\Vert v_{1}\Vert\Vert v_{2}\Vert\Vert\varphi_{T}%
^{\ast}-\varphi_{0}\Vert\Tilde{\tau}_{n(T)}.
\]
Since $\varphi_{T}^{\ast}\overset{p}{\rightarrow}\varphi_{0}$ and $\tilde
{\tau}_{n(T)}\overset{p}{\rightarrow}\tilde{\tau}<\infty$ it follows from
(\ref{in app: random information}) that the right hand side tends in
probability to $0$. Hence $A_{T}(\hat{\varphi}_{T})\overset{p}{\rightarrow}0$
and \emph{(iii)} follows by (\ref{new: eq7}) using (\ref{in app: random score}).

\subsection{Proof of Theorem \ref{thm main}}

We apply Lemma \ref{new: main lemma} and verify that conditions (C.1)--(C.4)
hold with the counting process $n\left(  t\right)  $ defined by $n\left(
t\right)  =\max_{k}\{\sum_{i=1}^{k}x_{i}\leq t\}$, $t\in\lbrack0,\infty)$.

Condition (C.1) holds by Lee and Hansen (1994, Lemma 9). Next, under (i) and
(ii), by Meitz and Saikkonen (2008, Theorem 1), $v_{i}=\left(  x_{i},\psi
_{i}\right)  ^{\prime}$ is (geometrically) ergodic with $\mathbb{E}\left[
\left\Vert v_{i}\right\Vert \right]  <\infty$. Hence the strong LLN in Jensen
and Rahbek (2007) applies, and (C.2) and (C.3) hold by this and using the
arguments in Jensen and Rahbek (2004).

To verify (C.4), note that $\mathbb{E}\left[  x_{i}\right]  =\mu=\omega
_{0}/\left(  1-\alpha_{0}-\beta_{0}\right)  ,$ $0<\mu<\infty$, and $n^{-1}%
\sum_{i=1}^{n}x_{i}\overset{\text{a.s.}}{\rightarrow}\mu$ by the strong LLN in
Jensen and Rahbek (2007). Next, by Gut (2009, Theorem 2.1),
\[
n(T)^{-1}\sum_{i=1}^{n(T)}x_{i}\overset{\text{a.s.}}{\rightarrow}\mu,
\]
using $n(T)\overset{\text{a.s.}}{\rightarrow}\infty$ as $T\rightarrow\infty$.
As $T<\sum_{i=1}^{n(T)}x_{i}+x_{n(T)+1}\leq T+x_{n(T)+1}$, we get
\[
0\leq T/n(T)-\sum_{i=1}^{n(T)}x_{i}/n(T)<x_{n(T)+1}/n(T).
\]
Next, using $x_{n+1}/n\overset{\text{a.s.}}{\rightarrow}0$, Gut (2009, Theorem
2.1) implies that $x_{n(T)+1}/n(T)\overset{\text{a.s.}}{\rightarrow}0$, and
hence $n(T)/T\overset{\text{a.s.}}{\rightarrow}c=1/\mu$ as desired. That
$x_{n+1}/n\overset{\text{a.s.}}{\rightarrow}0$, follows from $n^{-1}\sum
_{i=1}^{n}x_{i}\overset{\text{a.s.}}{\rightarrow}\mu$, see e.g. Phillips and
Solo (1992, p.989) \hfill$\square$

\end{document}